\documentclass[conference]{IEEEtran}
\IEEEoverridecommandlockouts
\usepackage{cite}
\usepackage{amsmath,amssymb,amsfonts}
\usepackage{algorithmic}
\usepackage{graphicx}
\usepackage{textcomp}
\usepackage{xcolor}
\usepackage[hidelinks]{hyperref}

\def\BibTeX{{\rm B\kern-.05em{\sc i\kern-.025em b}\kern-.08em
    T\kern-.1667em\lower.7ex\hbox{E}\kern-.125emX}}
\begin{document}

\title{Type Systems in Resource-Aware Programming: Opportunities and Challenges}

\author{\IEEEauthorblockN{Alcides Fonseca}
\IEEEauthorblockA{LASIGE\\ Faculdade de Ciências da Universidade de Lisboa \\
Lisboa, Portugal \\
alcides@ciencias.ulisboa.pt \\
0000-0002-0879-4015}
\and
\IEEEauthorblockN{Guilherme Espada}
\IEEEauthorblockA{LASIGE\\ Faculdade de Ciências da Universidade de Lisboa \\
Lisboa, Portugal \\
gjespada@ciencias.ulisboa.pt \\
0000-0001-8128-7397}
}

\maketitle

\begin{abstract}
Type systems provide software developers immediate feedback about a subset of correctness properties of their programs. 
IDE integrations often take advantage of type systems to present errors, suggest completions and even improve navigation.
On the other hand, understanding the time and energy consumption of the execution of a program requires manual testing.

In this paper, we identify existing work on using type systems for energy awareness, and define the requirements for a practical approach, which the existing approaches do not address fully.

 Furthermore, we also discuss how existing type systems can help generalize refactors for energy-efficiency.
\end{abstract}


\section{Introduction}

The Karlskrona manifesto has raised awareness for the sustainability of the Software Development Process~\cite{Becker2015}. Developers should be conscious of the sustainable impact of their decisions at every stage of the software development lifecycle. One major concern regarding sustainability is energy-consumption.

It is expected that by 2025 computing and communication will account for 20\% of the global energy usage~\cite{DBLP:journals/software/FonsecaKL19}. Energy efficiency is important for users both at the macro and micro scale. Data-centers account for 3\% of global emissions and represent a substantial running cost for online service providers. When evaluating new server purchases, the monthly energy cost is nowadays more important than the equipment's upfront cost. On the micro-scale, users are concerned about the battery life of their computers, tablets, phones and wearables.

The Manifesto for Energy-Aware Software~\cite{DBLP:journals/software/FonsecaKL19} contends that software developers must become energy-aware during all phases of software development. Furthermore, developers should be equipped with the necessary training and tools.

A thorough survey on the status of energy efficiency in software development~\cite{Pinto2017} concludes that most performance estimation tools focus on low-level rather than high-level programming at which application developers work. Software refactoring techniques and tools to detect and automatically fix energy inefficiencies with a direct impact on energy-consumption have been proposed in the past. However, these tools apply mostly to mobile application development, an area that directly impacts end users. The survey also concludes that there is a clear lack of knowledge in writing, maintaining, and evolving energy-efficient software systems.

\section{Static Analysis for Energy-awareness}

SEEP~\cite{DBLP:journals/sigops/HonigEKS11} is a framework that uses platform-specific energy profiles and symbolic execution to reason about the energy consumption of a full program, using a compositional approach on top of the low-level profiling. This approach has shown to have small errors, in the order of ~0.09 mJ. Moreover, static analysis has scaled to detect energy bugs in Android applications~\cite{DBLP:conf/icfem/JiangYQSZY17}.

Different static analysis tools have been used: abstract interpretation~\cite{DBLP:journals/corr/Lopez-GarciaHKL15}, relationship analysis~\cite{DBLP:conf/icfem/JiangYQSZY17}, and Machine-Learning~\cite{DBLP:journals/sncs/MarantosSPS21}.

A review of static analysis for energy consumption prediction~\cite{DBLP:journals/corr/GeorgiouKE15} identifies the need for these approaches to be aware of data distributions. Most approaches work towards the upper bound, but developers often want to understand the behavior in average cases.

\section{Type Systems for Energy-awareness}

The evaluation of energy efficiency is traditionally done either in simulation or in testing, due to the necessity to have an accurate measure. This creates a disconnect between writing code and understanding its energy efficiency. It has been shown that immediate feedback is beneficial for correctness properties (e.g., using contracts~\cite{DBLP:journals/scp/LozanoMK15}), but the advantages of immediate feedback for non-functional properties has been less studied.

We argue that a more immediate feedback would help developers learn about what in their code consumes more energy and should be used more carefully. It is also important for energy-awareness education~\cite{DBLP:journals/software/FonsecaKL19}.

Type systems have succeeded in improving developer productivity, detecting correctness problems early in the development process. These tools are integrated in the IDE to provide immediate feedback to prevent users from writing incorrect code. Otherwise, developers would only detect those errors at runtime when (or if) that code path is executed, as it is common in dynamically-typed languages. And even these languages have been adopting gradual typing (e.g., Typescript and Python Type annotations) to take advantage of this ergonomic advantage.

Based on our previous work on energy efficiency~\cite{DBLP:conf/sblp/MelfeFF18,DBLP:conf/greens/MelfeFF18,DBLP:journals/tsusc/FonsecaCCB18,DBLP:journals/suscom/FonsecaC18,DBLP:journals/pc/FonsecaC18}, we identify the following requirements for Energy-aware programming:

\begin{enumerate}
	\item Feedback must be instantaneous — For this system to be useful, developers should get immediate feedback without having to run the full program.
	\item The resource cost should be symbolic — Because the cost of the program may depend on arguments, the present cost should reflect that dependency. If arguments are passed, then the absolute value should be considered.
	\item The resource usage follows a probabilistic model — Running the same program twice does not take the same amount of time or energy. As such, users may want to inspect the distribution of values. Users may want to reason about median values, or the 95\% percentile, as it is common in Service Level Agreements (SLA).
\end{enumerate}

To address these requirements, we are developing a modular approach centered on the cost of functions. The cost of a function can be inferred by what it performs, with standard library low-level functions annotated based on micro-benchmarks.

To compose the cost of functions, we intend to follow existing work on verification of resource usage~\cite{Cohen2012,Zhu2015,Gastel2015,Canino2017}. However, these approaches are not sufficient to deal with the energy-consumption indirection between high-level code and the low-level code that is executed. In particular, we identify the following as major features that need to be accounted for to achieve a reasonable accuracy: 

\begin{enumerate}
	\item Garbage Collection;
	\item Just-in-Time Compilation;
	\item Operating System Scheduler Overheads.
\end{enumerate}

Kersten et al. presented a hardware-parametrized Hoare logic for energy consumption analysis to verify if a program has a correct upper bound of energy consumption~\cite{Kersten2013}. But this approach does consider the previously mentioned features, nor the probabilistic approach.

Gastel, Kersten and Eekelen present an approach of using dependent types to define the energy semantics~\cite{Gastel2015}. This work is the closest to our approach because a type system is used to keep track of energy consumption. The major advantages are its modularity and precision (compared with the bounded approach of Hoare logic). However, this approach also does not consider the probabilistic nature of executions, as well as the features that introduce indirection.

\section{Type-based refactorings for Energy-awareness}

As the scientific community has been better understanding the performance impact of design decisions, new methods for improving energy-efficiency have been proposed. Cruz and Abreu proposed a limited set of automatic refactorings for improving the energy efficiency of Android applications that range from moving allocations to changing UI elements~\cite{Cruz2017}. Similarly, Morales et al. analyzed 8 anti-patterns in Android applications, presented refactorings within a minute, and extended the mobile phone's battery life by up to 29 minutes. Developers found that 68\% of the suggested refactorings to be very relevant~\cite{Morales2018}.

Sehgal et al. generalize this energy-efficiency refactoring approach to general-purpose software, relying on readings from different components to evaluate each refactor~\cite{Sehgal2020}. Kim et al. applied the same approach for embedded computing environments to improve legacy code performance~\cite{Kim2018}. Imran and Kosar evaluated auto-refactorings in the context of Cloud Software with significant improvements in energy consumption.

Most of the existing approaches have focused on developing tools to reason about existing applications' energy-efficiency and suggest improvements. Energy Types is a first effort that allows annotating a Java-like language with energy-related modes that can be used to reason about the resource consumption of apps and allow the compiler to introduce dynamic voltage and frequency control statements in the generated binary, selecting the best configuration for different parts of the code~\cite{Cohen2012}. The authors admit that this approach is neither the most precise in characterizing energy consumption nor the most expressive in capturing energy-aware programmers' intention. ECO applies this technique to Java programs and evaluates this approach using energy usage~\cite{Zhu2015}. ENT is another extension of Java supporting both static and dynamic types to control and reason about energy-consumption~\cite{Canino2017}. Ent allows developers to control energy usage (static types) or how to react to a given change (dynamic type). Ent helps the developer in debugging and identifying energy hotspots in their applications.

As such, there is a gap between refactorings (based on particular domains) and the type systems designed to reason about code. We identify bridging this two gaps as an opportunity to generalize more refactors that are based on the energy-consumption patterns of general-purpose code.

\section{Conclusions}

There have been some advances on modeling energy consumption at the code level, but there are many factors that are not being considered (Garbage Collection, Just-in-time Compilation, Memory Allocation). Additionally, practitioners often define service level agreements based on percentiles, not by averages or upper bounds as one would be optimistic and the other would be impossible pessimistic.

Static analysis has been successful in estimating resource consumption, and Liquid and Dependent Types are capable of doing abstract interpretation, a popular technique for low-level estimation of energy consumption of programs.
Furthermore, to answer the need of reasoning about energy consumption in a statistical manner, instead of considering only worst-case scenarios or average cases, probabilistic programming languages can be useful to improve type systems for energy consumption.

\section*{Acknowledgment}

This work was supported by the Fundação para a Ciência e a Tecnologia (FCT) under LASIGE Research Unit (UIDB/00408/2020 and UIDP/00408/2020). Experimental evaluation was conducted on hardware provided by projects CMU|Portugal CAMELOT (LISBOA-01-0247-FEDER-045915) and Resource-Aware Programming (EXPL/CCI-COM/1306/2021) .

\bibliographystyle{IEEEtran}
\bibliography{bibliography}
\end{document}